\documentclass[12pt,preprint]{aastex}
\begin{document}

\title{What Is The Neon Abundance Of The Sun?}
\author{John N. Bahcall}
 \affil{Institute for Advanced Study, Einstein Drive,Princeton, NJ 08540}
\author{Sarbani Basu}
\affil{Department of Astronomy, Yale University, New Haven, CT 06520-8101}
\and
\author{Aldo  M. Serenelli}
 \affil{Institute for Advanced Study, Einstein Drive,Princeton, NJ 08540}
\begin{abstract}
We have evolved a series of thirteen complete solar models that
utilize different assumed heavy element compositions. Models that
are based upon the heavy element abundances recently determined by
Asplund et al. (2005)\nocite{asplundgrevessesauval2005} are inconsistent
with helioseismological measurements. However, models in which the
neon abundance is increased by 0.4-0.5 dex to $\log N({\rm Ne})  =
8.29 \pm 0.05$ (on the scale in which $\log N({\rm H})  = 12$) are
consistent with the helioseismological measurements even though
the other heavy element abundances are in agreement with the
determinations of Asplund et al.
(2005)\nocite{asplundgrevessesauval2005}. These results sharpen
and strengthen an earlier study by Antia \& Basu
(2005)\nocite{antia05}. The predicted solar neutrino fluxes are
affected by the uncertainties in the composition by less than
their $1\sigma$ theoretical uncertainties.
\end{abstract}

\section{INTRODUCTION}
\label{sec:introduction}

Solar models constructed with recently determined heavy element
abundances (Asplund et al. 2005; see also Lodders
2003)\nocite{asplundgrevessesauval2005}\nocite{lodders03} disagree by many times the
quoted measuring errors with helioseismological determinations of
solar properties (Bahcall et al. 2005b)\nocite{BS05}.
This disagreement has led to a number of attempts to reconcile the
solar models and the helioseismological measurements. So far there
has not been a successful resolution of this problem (see, for
example, Bahcall \& Pinsonneault 2004;\nocite{BP04} Bahcall et
al. 2004\nocite{BSP04}; Bahcall et al. 2005b;\nocite{BS05} Basu \& Antia 2004;\nocite{basu04}
Antia \& Basu 2005;\nocite{antia05} Turck-Chi\`eze et al.
2004;\nocite{TC} Guzik \& Watson 2004; \nocite{guzikwatson04}
Guzik et al. 2005; Seaton \& Badnell 2004;\nocite{seaton}
Badnell et al. 2005; Montalban et al.
2004).\nocite{badnelletal2004}\nocite{montalban}

Revising the radiative opacity used in the solar models appeared
initially to be the most promising avenue for reaching agreement
(Bahcall et al. 2004,\nocite{BSP04} Basu \&
Antia 2004,\nocite{basu04} Bahcall et al. 2005a).\nocite{BBPS05} However, a recalculation and
refinement by the Opacity Project (OP, Badnell et al.
2005\nocite{badnelletal2004}) of the radiative opacity used in the
solar models yielded values for the radiative opacity that were
very close to the previously used values (OPAL, Iglesias\& Rogers
1996)\nocite{opalopacity96}, and therefore the revision is
insufficient to explain the disagreement between solar modelling
and helioseismology.  Of course, there could still be inadequacies
in the calculation of the opacity in the region of interest that
are common to both OPAL and OP evaluations.

It is also natural to ask if changing the theoretical diffusion rate
from the standard estimates could remove the discrepancies.  Guzik et
al. (2005)\nocite{guzikwatsoncox05} have explored solar models with a
variety of diffusion treatments and conclude that the required
increases in diffusion rates are unphysically large (for example, an
order of magnitude increase compared to the 15\% uncertainty estimated
by Thoul et al. 1994)\nocite{thoul} (see also Turcotte et
al. 1998;\nocite{turcotteetal98} Delahaye \& Pinsonneault
2005;\nocite{delahaye} and Seaton
2005)\nocite{seatonradiative}. Moreover, none of the variations of the
diffusion rates restored completely the good agreement between
helioseismology and solar modelling that was obtained (e.g., Bahcall
et al.  2005a\nocite{BBPS05}) with the previously-estimated (higher)
heavy element abundances of Grevesse \& Sauval (1998).\nocite{oldcomp}

We note that the excellent agreement between the helioseismological
measurements and the solar model predictions made using the older
Grevesse \& Sauval (1998)\nocite{oldcomp} abundances could be an
accident.  There are many uncertainties in the treatment of the
external layers of the star. The discrepancies between solar model
predictions made with the newer Asplund et
al. (2005)\nocite{asplundgrevessesauval2005} abundances and
helioseismological measurements represent a stimulating challenge to
make further improvements.

Could the discrepancies between helioseismology and solar
modelling be due to errors in one (or a few) poorly measured
element abundances? Antia \& Basu (2005)\nocite{antia05} have
constructed envelope models of the Sun which incorporate the
seismologically-determined profile of sound speeds and the
seismologically-inferred abundance of hydrogen and depth of the
convective zone.  They constructed a series of envelope models in
which they changed the abundance of certain elements in order to
test the effect of the abundance changes on the
helioseismologically determined density profile. Antia \& Basu
(2005)\nocite{antia05} inferred that (with OP opacities) an
increase in the Ne abundance by $0.63 \pm 0.06$ dex relative to
the Asplund et al. (2005)\nocite{asplundgrevessesauval2005}
preferred value yielded agreement with the density profile. Thus
they found that, with the assumptions made regarding the envelope
model, an abundance of [Ne/H] = 8.47 reconciled the disagreement
between envelope models and helioseismology. Moreover, Antia and
Basu found that satisfactory agreement with the density profile
could also be achieved by simultaneously increasing the C, N, O,
and Fe abundances by 0.05 dex and the Ne by 0.40 dex relative to
the Asplund et al. (2005)\nocite{asplundgrevessesauval2005}
recommended values.

In this paper, we evolve  a series of complete solar models,
interior plus atmosphere, to test the effects of abundance changes
upon the full set of helioseismologically determined parameters:
sound speed profile, depth of the convective zone, surface helium
abundance, and density profile.  The complete solar models have
three advantages over envelope models: 1) the complete models
incorporate solar evolution and a full description of the solar
interior; 2) the complete models can be tested against all of the
helioseismologically determined parameters, not just the density
profile; and 3) the complete models can be used to determine how a
given composition affects the calculated solar neutrino fluxes.

Like Antia and Basu, we consider that the neon abundance could
possibly change by much more than the quoted uncertainty, 0.06 dex
(Asplund et al. 2005).\nocite{asplundgrevessesauval2005} The
reason is that neon can not be measured spectroscopically in the
solar photosphere and must be determined in regions in which the
physical conditions are less well understood (see, e.g., the
discussion in Lodders 2003)\nocite{lodders03}. For exactly the
same reason, we consider that the argon abundance could possibly
change by much more than the quoted uncertainty, 0.08 dex. In
addition, it should be noted that the contribution of neon to the
radiative opacity in the solar radiative interior is very
important in the range of temperatures between $2-5\times10^6$~K,
precisely in the region the opacity needs to be increased in order
to solve the discrepancy between solar model predictions and
helioseismology (Bahcall et al.
2005a).\nocite{BBPS05}

 We present in
\S\ref{sec:modelsdifferentcompositions} and Table~\ref{tab:models}
our results for a series of thirteen complete solar models with
different chemical compositions and in \S\ref{sec:fluxes} and
Table~\ref{tab:fluxes} we give the solar neutrino fluxes for each
of these models. We describe our main results and discuss our
conclusions in \S\ref{sec:discussion}.

\section{SOLAR MODELS WITH DIFFERENT COMPOSITIONS}
\label{sec:modelsdifferentcompositions}

\begin{deluxetable}{ccccccccccc}
\tabletypesize{\scriptsize}
\tablecaption{Solar Models with Different Compositions.
\label{tab:models}}
\tablehead{
\colhead{Model} & \colhead{Ne} &  \colhead{Ar} &  \colhead{CNO} &
\colhead{Si+}
  & \colhead{$Z/X$}  & \colhead{$Z_{\rm i}$}  & \colhead{$Y_{\rm
Surf}$} &
\colhead{$R_{\rm BC}$} & \colhead{$\sqrt{<\delta c^2/c^2>}$}
& \colhead{$\sqrt{<\delta \rho^2 / \rho^2>}$}}
\startdata
BS05(OP) &  --- & ---  & ---  & --- &  0.02292 &
0.01884  & 0.2426 &  0.7138 &
0.0009 & 0.012 \\
&&&&&&& (1.7) & (0.5) & & \\

BS05(AGS,OP) & --- &  --- & --- & --- & 0.01655 & 0.01404  &
0.2292 & 0.7281 &
0.0049 & 0.045 \\
&&&&&&& (5.7) & (15.) & & \\

\tableline Model~3 & 0.45 & 0.4 & 0.05 & 0.02 & 0.02069 & 0.01700 &
0.2439 &
0.7146 & 0.0010 & 0.011 \\
&&&&&&& (1.4) & (1.3) & & \\

Model~4 & 0.4 & 0.4 & 0.05 & 0.02 & 0.02027 & 0.01671 & 0.2425 &
0.7163 & 0.0012 & 0.015\\
&&&&&&& (1.8) & (3.0) & & \\

Model~5 & 0.4 & --- & 0.05 & 0.02 & 0.02018 & 0.01667 & 0.2404 &
0.7153 & 0.0011 &
0.014\\
&&&&&&& (2.4) & (2.0) & & \\

Model~6 & 0.4 & 0.4  & 0.05 & --- & 0.02008 & 0.01662 &  0.2400 &
0.7164 & 0.0013 &
0.018 \\
&&&&&&& (2.5) & (3.1) & & \\

Model~7 & 0.4 & ---  & 0.05 & --- & 0.01998 & 0.01657 &  0.2378 &
0.7155 & 0.0011 &
0.016 \\
&&&&&&& (3.1) & (2.2) & & \\

Model~8 & 0.45  & 0.4 & --- & --- &  0.01916 & 0.01584 & 0.2411 &
0.7174 & 0.0013 &
0.016 \\
&&&&&&& (2.2) & (4.1) & & \\

Model~9 & 0.4  & --- & --- & --- &  0.01864 & 0.01550 & 0.2374 &
0.7183 & 0.0015 &
0.018 \\
&&&&&&& (3.3) & (5.0) & & \\

Model~10 & 0.5  & --- & --- & --- &  0.01954 & 0.01614 & 0.2405 &
0.7150 & 0.0012 &
0.009 \\
&&&&&&& (2.4) & (1.7) & & \\

Model~11 & 0.6  & --- & --- & --- &  0.02068 & 0.01694 & 0.2440 &
0.7116 & 0.0020 &
0.007\\
&&&&&&& (1.3) & (1.7) & & \\

Model~12 & 0.6  & 0.4 & 0.05& 0.02 &  0.02230& 0.01814 & 0.2486 &
0.7099& 0.0021 &
0.010\\
&&&&&&& (0.0) & (3.4) & & \\

Model~13 & 0.3  & 0.4& 0.05 & 0.02 &  0.01955 & 0.01621 & 0.2402 &
0.7187 & 0.0019 &
0.022 \\
&&&&&&& (2.4) & (5.4) & & \\
\enddata
\tablecomments{The
first two rows of the table summarize the predicted properties for
a standard solar model, BS05(OP), that assumes the Grevesse \&
Sauval (1998) heavy element abundances and a standard solar model,
BS05(AGS, OP) that the assumes the Asplund et al. (2005) heavy
element abundances. Both models were evolved by Bahcall, Serenelli
\& Basu (2005) using the recently-computed Opacity Project
(Badnell et al. 2004) radiative opacities. The solar model
quantities given in the table are: the total surface value of the
mass of heavy elements relative to hydrogen ($Z/X$), the initial
heavy element abundance ($Z_i$), the present-day surface helium
abundance ($Y_{\rm surf}$), the depth of the base of the
convective zone ($R_{cz}$, in units of $R_{\odot})$, and the rms
fractional differences between the model and the solar sound
speeds ($\sqrt{<\delta c^2/c^2>}$) and the model and the solar
density distribution($\sqrt{<\delta \rho^2 / \rho^2>}$). The
numbers in parentheses are the differences between the measured
values given in equation~(\ref{eq:convectivezone}) and
equation~(\ref{eq:helium}) and the calculated solar model values, in
both cases divided by the relevant estimated measuring error. The
last nine rows of the table present the parameters that are
predicted by solar models with compositions that differ from the
Asplund et al. (2005) compositions by adding different amounts of
neon, argon, CNO, and elements for which the meteoritic abundances
are well determined (which we abbreviate by Si+). For example, the
third row of the table gives the results obtained with a solar
model that has the Asplund et al. (2005) solar composition
supplemented by 0.45 dex of neon, 0.4 dex of argon, 0.05 dex for
carbon, nitrogen, and oxygen, and 0.02 dex for elements for which
the meteoritic abundance is well measured.}
\end{deluxetable}

In this section, we compare the results obtained from a series of
solar models that have  different assumed heavy element abundances
with the depth of the solar convective zone, the surface helium
composition, and the sound speed and density distributions, all
obtained from helioseismological measurements.

The depth of the convective zone that is inferred from
helioseismological measurements is (Basu \& Antia
2004).\nocite{basu04}
\begin{equation}
\frac{R_{\rm CZ}}{R_\odot} ~ = ~ 0.7133 \pm 0.001,\, {\rm
~measurement}. \label{eq:convectivezone}
\end{equation}

The inferred surface abundance of helium is (Basu \& Antia
2004).\nocite{basu04}
\begin{equation}
 Y_{\rm surf} ~ = ~ 0.2485 \pm 0.0034,\, {\rm
~measurement}\, .
\label{eq:helium}
\end{equation}

The sound speed and the density can be measured at a number of
different radial depths in the Sun. Fortunately, the inferred
distributions of sound speeds and densities are, to an excellent
approximation, independent of the reference model used to derive the
solar distributions.  For example, making different choices for the
reference model causes variations of only of order 0.03\% in the
profile of the sound speeds and of order 0.3\% in the density profile
(Basu et al. 2000).\nocite{BBP00} These changes
are small compared to the variations in the profiles predicted by
different solar models that have significantly altered surface
abundances. Therefore, we compare all of the profiles reported in this
section with the solar sound speed and density profiles inferred with
the reference model BS(AGS, OP), which uses the Asplund et
al. (2005)\nocite{asplundgrevessesauval2005} heavy element abundances
and the recently-calculated (Badnell et
al. 2004)\nocite{badnelletal2004} radiative opacities.

Table~\ref{tab:models} shows the calculated solar properties for a
series of theoretical solar models that were evolved with different
assumed heavy element abundances. The first two rows in the table
present results for models with the Grevesse \& Sauval
(1998)\nocite{oldcomp} abundances (row~1) and the Asplund et al.
(2005)\nocite{asplundgrevessesauval2005} abundances (row~2). In rows
3-13, we present the calculated solar quantities for solar models with
compositions that differ from the Asplund et al.  (2005) composition
primarily by the addition of neon and argon (with, in some cases, a
touch, 0.05 dex, of additional CNO elements and a dash, 0.02 dex, of
the heavier elements whose abundances are well determined by
meteoritic measurements, see Asplund et
al. 2005;\nocite{asplundgrevessesauval2005} Lodders
2003)\nocite{lodders03}.

\begin{figure*}
\includegraphics[angle=90,scale=.65]{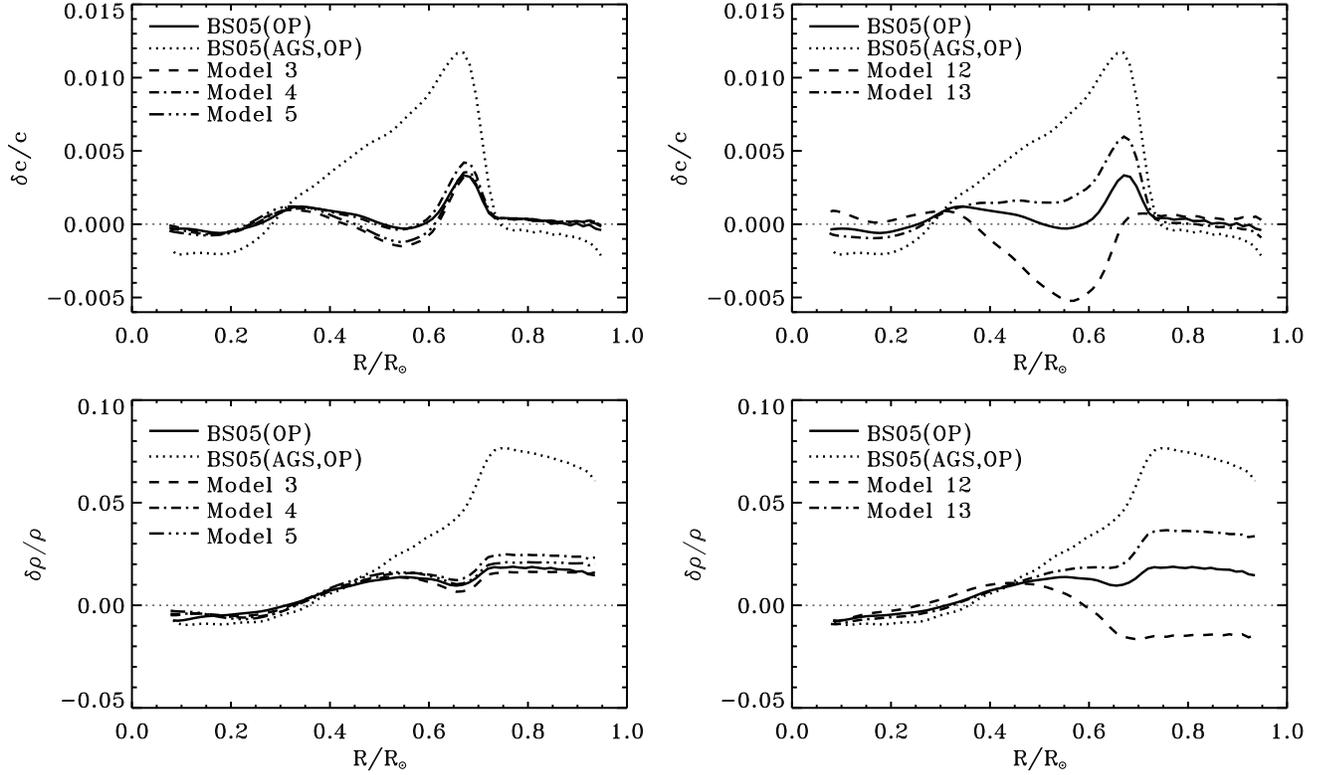}

\caption[]{Relative sound-speed differences, $\delta c/c=
(c_\odot-c_{\rm model})/c_\odot$ (top panels) and relative density
differences, $\delta \rho/\rho= (\rho_\odot-\rho_{\rm
model})/\rho_\odot$ (bottom panels) between some selected solar
models and helioseismological results from MDI data.  All panels
show the BS05(OP) and BS05(AGS,OP) standard solar models for
reference. In the two left panels, models 3, 4, and 5 (see
Table~\ref{tab:models} for details on the composition changes)
show an agreement with helioseismology data which is comparable to
that of our preferred standard solar model, BS05(OP). In the right
top panel, models 12 and 13 show a better agreement with the
helioseismological sound speed profile than the BS05(AGS,OP)
model, but are still a factor of two worse than BS05(OP).  Note in
particular the effect in the sound-speed of adding too much neon
(+0.6 dex, model 12). This is the result of an excessive
enhancement in the opacity, to which neon is a main contributor in
the solar interior between 0.4-0.7~R$_\odot$. Model~12 predicts,
on the contrary, an acceptably good density profile (a similar
result for the density profile was found by Antia \& Basu 2005
using envelope models). \label{fig:inversion}}

\end{figure*}

We have used trail-and-error in order to find a composition
mixture that fits well the helioseismological data without
exceeding the quoted uncertainty (Asplund et al.
2005)\nocite{asplundgrevessesauval2005} in any of the abundance
determinations except for neon and argon.

The model described in the third row of Table~\ref{tab:models} fits
very well all the helioseismological measurements.  In fact, this
model, which differs from the Asplund et
al. (2005)-based\nocite{asplundgrevessesauval2005} BS05(AGS, OP) model
primarily by adding significant amounts of neon and argon, fits the
helioseismological data essentially as well as the model that assumes
the element abundances of Grevesse \& Sauval
(1998)\nocite{oldcomp}. The results for the other models in
Table~\ref{tab:models} show that there are a variety of satisfactory
choices for the altered composition with the increase in the neon
abundance between 0.4 dex and 0.6 dex. An increase of only 0.3 dex
appears to be insufficient (see Model~13); this model has too shallow
a depth for the convective zone. An increase of as much as 0.6 dex
appears to be too much; the rms difference between the solar and the
model sound speed is more than twice as large as for the BS05(OP)
model.

Figure~\ref{fig:inversion} shows the fractional differences between
the sound speeds and densities calculated with selected solar models
and those inferred with the aid of the Michelson Doppler Imager (MDI)
on board the Solar and Heliospheric Observatory (SOHO). In particular,
we use frequencies obtained from MDI data that were collected for the
first 360 days of its observation (Schou et
al.~1998)\nocite{sch98}. For reference, we show in all four panels of
Figure~\ref{fig:inversion} the BS05(OP) solar model (model~1 of
Table~\ref{tab:models}, made using the older Grevesse \& Sauval
1998\nocite{oldcomp} higher heavy element abundances) and the
BS05(AGS, OP) solar model (model~2 of Table~\ref{tab:models}, made
using the Asplund et al. 2005 lower heavy element abundances). The
three other models that are illustrated in the left hand panels of
Figure~\ref{fig:inversion} (models 3-5 of Table~\ref{tab:models}) were
selected in order to illustrate that by a judicious choice of the neon
abundance it is possible to obtain agreement with the
helioseismological sound speed and density profiles that is comparable
to what is obtained with our preferred solar model, BS05(OP).

The top right panel shows that adding neon is not a panacea. The panel
establishes this result with the aid of two solar models from
Table~\ref{tab:models}, model~12 (+0.6 dex of neon respect to Asplund
et al. 2005)\nocite{asplundgrevessesauval2005} and 13 (+0.3 dex of
neon with respect to Asplund et al. 2005). Both models 12 and 13 yield
a factor of two worse agreement with the helioseismological sound
speed profile than BS05(OP) (although they are in better agreement
with the profile than BS05(AGS, OP), see also Table~\ref{tab:models}).

Model~12 is particularly instructive. The agreement with the
helioseismologically determined density profile is acceptable. A
similar result was found by Antia \& Basu (2005)\nocite{antia05}
in their study of envelope models. The Antia-Basu envelope models
could only be tested against the density profile since their
models had by construction the correct sound speed profile.
However, the predicted sound speed profile of Model~12 is much
worse than for BS05(OP)and Figure~\ref{fig:inversion} shows that
the sound speed profile is unacceptable. The principal reason that
the profile of the sound speed is unacceptable while the density
profile is acceptable is that the sound speed is measured much
more accurately than the density.

We note that the successful composition changes are essentially a way
of increasing the opacity in the region in which the model predictions
differ from the measured helioseismological properties. The required
corrections in the opacity must extend from $2 \times 10^6$K ($R =
0.7R_\odot$) to $5\times 10^6$K ($R = 0.4 R_\odot$), with opacity
increases of order 10\% (Bahcall et al. 2005a)\nocite{BBPS05}.

\section{NEUTRINO FLUXES FOR DIFFERENT ASSUMED COMPOSITIONS}
\label{sec:fluxes}

We present in this section the neutrino fluxes computed for each
of the  solar models that we have discussed in the previous
section. The main purpose of this discussion is to show that the
predicted neutrino fluxes are insensitive to the composition
changes that are being considered.

\begin{deluxetable}{ccccccccccccc}
\tablecaption{Neutrino fluxes for Solar Models with Different
Compositions.  \label{tab:fluxes}\medskip}
\tablewidth{0pt}
\tablehead{
\colhead{Model} &  \colhead{pp} & \colhead{pep}
& \colhead{hep} & \colhead{Be7} & \colhead{B8} & \colhead{N13}
& \colhead{O15} & \colhead{F17}
}
\startdata
BS05(OP) &  5.99 & 1.42 & 7.93 & 4.84 & 5.69 & 3.07
&
2.33 & 5.84 \\ \\
BS05(AGS,OP) & 6.06 & 1.45 & 8.25 & 4.34 & 4.51 & 2.01
& 1.45 & 3.25 \\ \\
\hline Model~3 &  6.00 &1.43 & 7.96 &
4.83 & 5.57 & 2.51 & 1.89 & 4.33 \\ \\

Model~4 &  6.01 &1.43 & 7.98 & 4.78
& 5.47 & 2.48 & 1.87 & 4.27 \\ \\

Model~5 & 6.02 & 1.44 & 8.03 & 4.70 & 5.30 & 2.44 &
1.82 & 4.15 \\ \\

Model~6 & 6.02 & 1.44 & 8.04 & 4.68 & 5.23 & 2.42 &
1.80 & 4.11 \\ \\

Model~7 & 6.03 & 1.44 & 8.09 & 4.60 & 5.06 & 2.38 &
1.76 & 3.99 \\ \\

Model~8 & 6.02 & 1.44 & 8.04 & 4.70 & 5.26 & 2.16 &
1.61 & 3.67 \\ \\

Model~9 & 6.03 & 1.45 & 8.11 & 4.58 & 4.99 & 2.10 &
1.55 & 3.52 \\ \\

Model~10 & 6.02 & 1.44 & 8.06 & 4.67 & 5.20 & 2.14 &
1.59 & 3.63 \\ \\

Model~11 &  6.01 & 1.44 & 7.99 & 4.79 & 5.46 & 2.20 &
1.65 & 3.78 \\ \\

Model~12 & 5.98 & 1.43 & 7.87 & 4.99 & 5.94 & 2.59 &
1.98 & 4.56 \\ \\

Model~13 & 6.01 & 1.43 & 8.03 & 4.70 & 5.30 & 2.44 & 1.83 &
4.16 \\
\enddata
\tablecomments{The table gives the neutrino fluxes computed for
the same models that are described in Table~\ref{tab:models}. The
 fluxes are presented in units of $10^{10}(pp)$, $10^{9}({\rm \, ^7Be})$,
$10^{8}({\rm pep}, {\rm ^{13}N, ^{15}O})$, $10^{6} ({\rm \, ^8B,
^{17}F})$, and $10^{3}({\rm hep})$ ${\rm
cm^{-2}s^{-1}}$.}
\end{deluxetable}

Table~\ref{tab:fluxes} presents the calculated solar neutrino
fluxes for all thirteen of the solar neutrino models that are
listed in Table~\ref{tab:models}.  The fluxes predicted by models
3-13 fall between the extremes for the first two models, BS05(OP)
(which assumes the Grevesse \& Sauval 1998\nocite{oldcomp}
abundances) and BS05(AGS, OP) (which assumes the Asplund et al.
2005 abundances)\nocite{asplundgrevessesauval2005}.

In the absence of a definitive reason for choosing between the
Grevesse \& Sauval (1998)\nocite{oldcomp} abundances and the
Asplund et al. (2005)\nocite{asplundgrevessesauval2005}
abundances, it is reasonable to consider for each flux the average
of the value predicted by BS05(OP) and the value predicted by
BS05(AGS,OP). For all the models listed in Table~\ref{tab:fluxes},
each of the neutrino fluxes differs by less than the $1\sigma$
theoretical uncertainty from the average flux predicted by the
BS05(OP) and BS05(AGS,OP) models, where the theoretical
uncertainties are given in Bahcall \& Serenelli
(2005)\nocite{derivatives}, Bahcall et al. (2005b)\nocite{BS05}.

We conclude that the variations in composition discussed in this
paper do not significantly affect the predicted solar neutrino
fluxes.

\section{DISCUSSION}
\label{sec:discussion}

Table~\ref{tab:models} shows that solar models with a variety of
chemical compositions, but all with an increase in [Ne/H] by 0.4-0.5
dex (a factor of 2.5 to 3.2) relative to the Asplund et
al. (2005)\nocite{asplundgrevessesauval2005} recommended value, yield
helioseismological parameters that are in satisfactory agreement with
all the helioseismological measurements. For example, BS05(OP) and
Models 3, 4, 5, and 10 all predict helioseismological parameters in
good agreement with what is measured. We are unable to find solar
models with an increase of neon abundance larger than 0.6 dex, or as
small as 0.3 dex, that are consistent with helioseismological (see,
e.g., Models 11, 12, and 13 of Table~\ref{tab:models}).

Thus we conclude that a neon abundance in the range [Ne/H] = 8.24
to 8.34 would, together with the Asplund et al.
(2005)\nocite{asplundgrevessesauval2005} values for other element
abundances (perhaps modified by their $1\sigma$ uncertainties),
resolve the conflict between helioseismology and solar model
predictions. This general result is in good agreement with the
inferences of Antia \& Basu (2005)\nocite{antia05} based on
envelope models.

The summary by Asplund et al.
(2005)\nocite{asplundgrevessesauval2005} gives a neon abundance of
$7.84 \pm 0.06$ based upon abundance ratios measured relative to
oxygen in the solar corona and with energetic particles. It is not
clear from their discussion what particular measurements were used
in this evaluation and how one could be sure that the abundance
ratio of neon to oxygen measured in the corona or with energetic
particles was the same as the photospheric abundance. Lodders
(2003)\nocite{lodders03} gives a particularly thorough discussion
of the neon abundance as measured by various techniques (see
especially Meyer 1989;\nocite{meyer89} Widing
1997;\nocite{widing97} Reames 1998;\nocite{reames98} Holweger
2001;\nocite{holweger2001} Martin-Hernandez
2003\nocite{Martinhernandez}). She obtains, based upon a
normalization to the photospheric oxygen abundance, a final result
of $7.87 \pm 0.10$, in good agreement with the value obtained by
Asplund et al.  (2005)\nocite{asplundgrevessesauval2005}. Feldman
\& Widing (2003) combined results derived from spectra emitted by
a low lying solar flare  with those derived from spectra emitted
by several freshly emerging high temperature plasmas regions  to
obtain a Ne/Mg photospheric abundance ratio of $3.38 \pm 0.20$.
Assuming a meteoritic determined abundance of 7.58 for magnesium,
Feldman \& Widing determined a neon abundance of $8.11 \pm 0.10$,
where the uncertainty does not reflect any systematic errors (see
especially Section~3 of Feldman \& Widing 2003)\nocite{feldman03}.

We see from the preceding paragraph that different methods yield
different values for the solar neon abundance.  One of the reason
for these discrepant values is that neon does not have any
suitable photospheric lines and therefore to obtain a
`photospheric' abundance for neon it is necessary to measure neon
relative to an element that does have good photospheric lines.
Different authors chose to normalize the neon abundance with
respect to different elements (e.g., Lodders normalizes relative
to oxygen and Feldman \& Widing normalize relative to magnesium).
In addition, all methods for determining the neon abundance are
uncertain because the environments (like the solar corona or upper
atmosphere) are imperfectly understood.

We have no idea whether the suggestion of a neon abundance in the
range [Ne/H] = $8.29 \pm 0.05$ dex is the correct solution to the
conundrum posed by the conflict between helioseismology and solar
models constructed with the Asplund et al.
(2005)\nocite{asplundgrevessesauval2005} abundances. Only new
observations with a variety of techniques and increased robustness
can decide this question.  The main reason for this paper to make
clear that further measurements of the neon abundance, accompanied
by careful analyses of the systematic uncertainties in the
measurements, are of great importance for solar physics.

If the correct solution to the conflict between solar modelling
and helioseismology is to increase the neon abundance $0.45 \pm
0.05$ dex, then this will be the first determination of a heavy
element abundance via helioseismology.  Of course, this
determination depends upon the assumption that there are no
relevant and important errors in other input parameters to the
solar model calculations and that we have not left out, as a
result of an unjustified approximation, a significant physical
process.

The recent investigations of solar abundances (Asplund
2000;\nocite{asplund00} Allende et al. 2001,
2002;\nocite{allende01,allende02} Asplund et
al. 2004)\nocite{asplund04} employ more powerful techniques than
previous used (see Grevesse \& Sauval 1998)\nocite{oldcomp} and
therefore command respect. However, it would be very valuable if
different groups would analyze the solar abundances using different
computer codes and different data.  The comparison between the results
of different groups would permit a more robust estimation of the
systematic uncertainties.

Table~\ref{tab:fluxes} shows that the neutrino fluxes for all
thirteen of the solar models considered in this paper lie within
$\pm 1\sigma$ theoretical uncertainties of each other, where the
theoretical uncertainties are summarized in Table~8 of Bahcall and
Serenelli (2005).\nocite{derivatives} Thus the uncertainty
regarding the abundances of heavy elements does not affect in an
important way the predicted solar neutrino fluxes.

After this paper was posted on the astrophysics archive but before
it was submitted to the Astrophysical Journal, we received a copy
of a very interesting paper by Drake \& Testa
(2005)\nocite{drake05}. These authors determine the neon to oxygen
abundance ratio from x-ray measurements of coronal lines in 21
nearby stars (median distance $\sim 30$ pc) with the Chandra X-ray
Observatory.  The x-ray observations indicate a neon to oxygen
ratio 2.7 times larger than the Asplund et al. (2005) value.  The
larger value from the x-ray observations is in good agreement with
the increase by a factor of $2.8 \pm 0.4$  that we require to fit
the helioseismological data. Drake and Testa also cite supporting
data from other x-ray and gamma-ray observations.

J. N. B. and A. M. S. are supported in part by NSF grant
PHY0070928 to the Institute for Advanced Study. S. B. is partially
supported by NSF grants ATM 0206130 and ATM 0348837. J. N. B.
thanks Marc Pinsonneault for repeatedly pointing out over the past
decade the potential importance and possible uncertainty of the
solar neon abundance.

\end{document}